\documentclass[a4paper]{jpconf}

\usepackage{graphicx}
\usepackage{acronym}
\usepackage{xcolor}
\usepackage{hyperref}
\usepackage{todonotes}
\usepackage{xspace}
\usepackage{amssymb}
\usepackage{amsmath}
\usepackage{bigfoot}

\usepackage[
    hyperref=true,
    url=false,
    isbn=false,
    backref=false,
    style=numeric,
    block=none,
    doi=true,
    eprint=false,
    sorting=none
]{biblatex}
\bibliography{BibTex/bibliography}

\hypersetup{
    colorlinks=true,
    linkcolor=black,
    filecolor=magenta,      
    urlcolor=blue,
    citecolor=blue,
}

\acrodef{ML}{machine learning}
\acrodef{PF}{particle-flow}
\acrodef{JetMET}{jets and missing energy}
\acrodef{DQM}{data quality monitoring}
\acrodef{GNN}{graph neural network}
\acrodef{LHC}{CERN Large Hadron Collider}
\acrodef{MLPF}{machine-learned particle-flow}
\acrodef{GCN}{graph convolutional network}
\acrodef{LSH}{locality sensitive hashing}
\acrodef{GPU}{graphics processing unit}
\acrodef{ECAL}{electromagnetic calorimeter}
\acrodef{HCAL}{hadron calorimeter}
\acrodef{PU}{pileup}
\acrodef{IPU}{intelligence processing unit}
\acrodef{MC}{Monte Carlo simulation}
\acrodef{CMSSW}{CMS software framework}
\acrodef{GSF}{Gaussian sum filter}
\acrodef{MC}{Monte Carlo}
\acrodef{HF}{hadron forward}
\acrodef{BREM}{Bremsstrahlung}
\acrodef{KF}{Kalman filter}
\acrodef{PUPPI}{pileup per particle identification}

\newcommand{\kt}{\ensuremath{k_{\mathrm{T}}}\xspace}
\newcommand{\ptmomentum}{\ensuremath{p_{\mathrm{T}}}\xspace}
\newcommand{\ptvecmiss}{\ensuremath{{\vec p}_{\mathrm{T}}^{\kern1pt\text{miss}}}\xspace}
\newcommand{\ptmiss}{\ensuremath{\ptmomentum^\text{miss}}\xspace}
\newcommand{\pthat}{\ensuremath{\hat{p}_\mathrm{T}}\xspace}

\newcommand{\ztautau}{\ensuremath{\PZ \rightarrow \PGt\PGt}\xspace}

\newcommand{\GEANTfour}{{\textsc{GEANT4}}\xspace}
\newcommand{\ONNXRUNTIME} {{\textsc{onnxruntime}}\xspace}

\newcommand{\akfourchs}[1]{AK4-CHS\xspace}
\newcommand{\akfourpuppi}[1]{AK4-PUPPI\xspace}
\newcommand{\GeV}{\ensuremath{\,\text{Ge\hspace{-.08em}V}}\xspace}
\newcommand{\TeV}{\ensuremath{\,\text{Te\hspace{-.08em}V}}\xspace}
\newcommand{\PGt}{\ensuremath{\mathrm{\tau}}\xspace}
\newcommand{\PZ}{\ensuremath{\mathrm{Z}}\xspace}
\newcommand{\Pe}{\ensuremath{\mathrm{e}}\xspace}
\newcommand{\Pgm}{\ensuremath{\mathrm{\mu}}\xspace}
\newcommand{\PGpz}{\ensuremath{\mathrm{\pi^0}}\xspace}
\newcommand{\PGppm}{\ensuremath{\mathrm{\pi^{\pm }}}}
\newcommand{\Pp}{\ensuremath{\mathrm{p}}\xspace}
\newcommand{\PGg}{\ensuremath{\mathrm{\gamma}}\xspace}
\newcommand{\Pn}{\ensuremath{\mathrm{n}}\xspace}

\newcommand{\ttbar}{\ensuremath{\mathrm{t\overline{t}}}\xspace}

\date{March 2023}

\begin{document}

\title{Progress towards an improved particle flow algorithm at CMS with machine learning}
\author{Farouk Mokhtar\textsuperscript{1}, Joosep Pata\textsuperscript{2}, Javier Duarte\textsuperscript{1}, Eric Wulff\textsuperscript{3}, Maurizio Pierini\textsuperscript{3} and Jean-Roch Vlimant\textsuperscript{4}\\
{\normalfont (on behalf of the CMS Collaboration)}}

\address{\textsuperscript{1}University of California San Diego, La Jolla, CA 92093, USA}
\address{\textsuperscript{2}NICPB, R\"{a}vala pst 10, 10143 Tallinn, Estonia}
\address{\textsuperscript{3}European Organization for Nuclear Research (CERN), CH 1211, Geneva 23, Switzerland}
\address{\textsuperscript{4}California Institute of Technology, Pasadena, CA 91125, USA}

\ead{fmokhtar@ucsd.edu, joosep.pata@cern.ch, jduarte@ucsd.edu}

\begin{abstract}
The \ac{PF} algorithm, which infers particles based on tracks and calorimeter clusters, is of central importance to event reconstruction in the CMS experiment at the CERN LHC, and has been a focus of development in light of planned Phase-2 running conditions with an increased pileup and detector granularity.
In recent years, the \ac{MLPF} algorithm, a \acl{GNN} that performs \ac{PF} reconstruction, has been explored in CMS, with the possible advantages of directly optimizing for the physical quantities of interest, being highly reconfigurable to new conditions, and being a natural fit for deployment to heterogeneous accelerators.
We discuss progress in CMS towards an improved implementation of the \ac{MLPF} reconstruction, now optimized using generator/simulation-level particle information as the target for the first time.
This paves the way to potentially improving the detector response in terms of physical quantities of interest.
We describe the simulation-based training target, progress and studies on event-based loss terms, details on the model hyperparameter tuning, as well as physics validation with respect to the current \ac{PF} algorithm in terms of high-level physical quantities such as the jet and missing transverse momentum resolutions.
We find that the \ac{MLPF} algorithm, trained on a generator/simulator level particle information for the first time, results in broadly compatible particle and jet reconstruction performance with the baseline \ac{PF}, setting the stage for improving the physics performance by additional training statistics and model tuning.
\end{abstract}
\acresetall
\section{Introduction}

\begin{figure}[ht]
    \centering
    \includegraphics[width=0.9\linewidth]{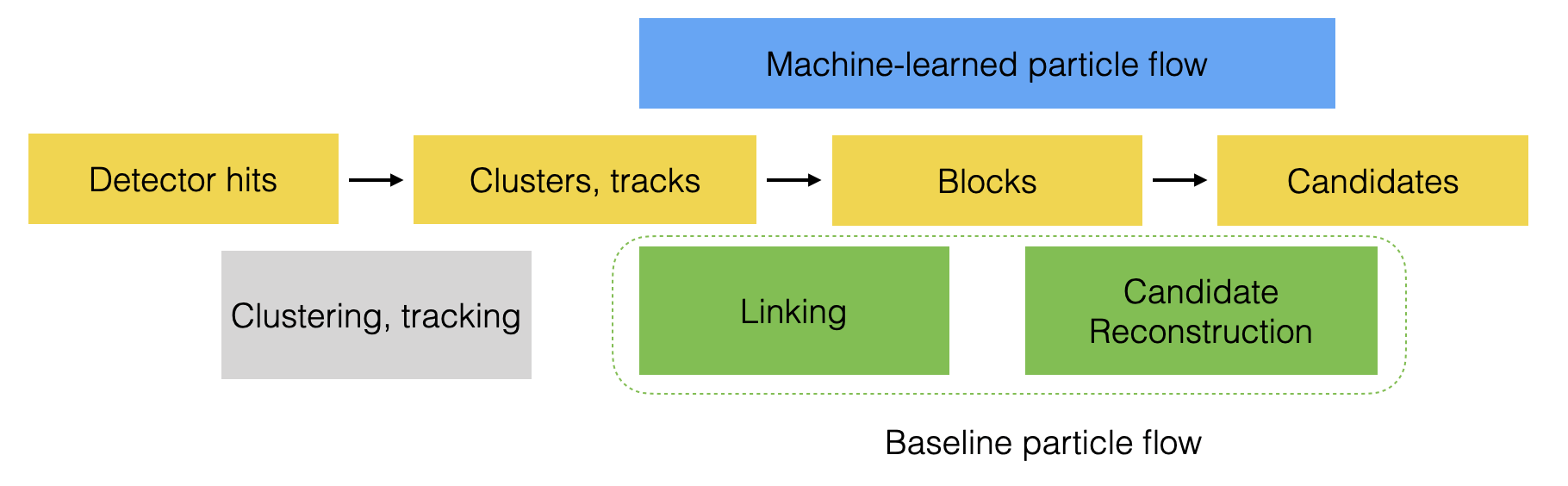}
    \caption{A schematic overview of \ac{MLPF} in the proposed reconstruction scheme.}
    \label{fig:mlpf}
\end{figure}

\Ac{PF} reconstruction is a global event reconstruction that combines information from different subdetectors in CMS (e.g. the tracker and the electromagnetic and hadronic calorimeters) to reconstruct stable particles~\cite{CMS:2017yfk}. 
The \ac{MLPF} algorithm is a \ac{GNN} trained to perform \ac{PF} reconstruction via supervised \ac{ML}~\cite{Pata:2021oez,Pata:2022wam,CMS-DP-2021-030}.
As with the baseline rule-based \ac{PF}, the inputs to \ac{MLPF} are tracks and calorimeter clusters (see \autoref{fig:mlpf}), and the outputs are stable \ac{PF} candidate particles.
The advantages of \ac{MLPF} include the possibility of deployment on heterogeneous computing accelerators (e.g. GPUs) and reoptimizing the algorithm in light of new experimental conditions.

In this work, we summarize the latest developments of \ac{MLPF} in CMS, which includes the development of a new generator/simulation-level training target, without referencing an existing \ac{PF} algorithm.
It should be noted that this effort fits in the wider context of implementing reconstruction algorithms for current and future detectors, partly or fully using trainable, differentiable \ac{ML} \cite{Alves:2023dkv,Pantaleo:2023rop,CMS:2022txh,DiBello:2020bas,DiBello:2022iwf}.
Results from our experiments show that we can achieve results that are largely compatible, and in some cases better, than standard \ac{PF}.
The full set of results is available in the accompanying Detector Performance note~\cite{CMS-DP-2022-061}.

\section{Particle-level training target definition}
The \ac{MLPF} training target is based on detector simulation information to closely approximate the input the simulation receives from the generator.
The truth particles for MLPF are the root nodes of the \GEANTfour simulation tree, consisting of simulated particle decays and interactions with matter, defined via the following truth definition algorithm.

The truth algorithm takes as input the tree of \GEANTfour simulation particles, searches for the earliest particles whose children leave detectable hits in either the tracker or calorimeters\footnote{We use the CMSSW \texttt{CaloParticle} and \texttt{TrackingParticle} modules to traverse the simulation tree}.
Given these decay tree root particles, we first address double-counting by removing the particles that have overlapping \GEANTfour simulation track identifiers.

Now knowing the set of root simulation particles whose decay products in principle interact with the detector, we have to define which of those we wish to reconstruct as \acs{PF} particles, and with which granularity.
The simulation particles are cleaned as follows:
\begin{enumerate}
    \item Coalesce particle labels according to \ac{PF} granularity: any charged hadrons are assigned to a single charged hadron class, all neutral hadrons are assigned to a single neutral hadron class, etc.
    \item Geometrically overlapping particles that leave energy deposits only to the same calorimeter cluster are not reconstructable separately, and are thus merged, keeping the label of the highest-energy particle.
    \item Electrons or muons with $\ptmomentum < 1 \GeV$ are relabeled as charged or neutral hadrons, based on the deposited track and calorimeter energy, to approximate the behaviour of the baseline \ac{PF} algorithm.
    \item to mimic the response of baseline \ac{PF}, particles outside the tracker acceptance are labeled as HF hadronic or HF electromagnetic, depending on the energy deposits to the respective calorimeters
\end{enumerate}

The resulting set of simulated particles is denoted as the \acs{MLPF} ground truth.
Comparisons between PF and the MLPF truth are available in pages 5--7 of the Detector Performance note~\cite{CMS-DP-2022-061}.

\section{Datasets}
With the algorithm above, we generate datasets for \acs{MLPF} truth validation and subsequent model optimization.
We use the official CMS software\footnote{\verb|CMSSW_12_3_0_pre6|} for sample generation, simulation and baseline \ac{PF} reconstruction, with the center of mass energy $\sqrt{s}=14\TeV$ and running conditions corresponding to Run 3.
For the samples with pileup, we use a flat \ac{PU} profile with a Poisson distribution in the range 55--75 , mixed in from a high-statistics minimum bias dataset\footnote{\verb|/RelValMinBias_14TeV/CMSSW_12_2_0_pre2-122X_mcRun3_2021_realistic_v1_HighStat-v1/GEN-SIM|}.
The training and validation is carried out on a mixture of all the samples listed in \autoref{tab:samples}.

\begin{table}[ht]
    \centering
    \begin{tabular}{c|c|c}
        physics process & \acs{PU} configuration & \acs{MC} events \\
        \hline
        top quark-antiquark pairs (\ttbar) & flat 55--75 & 100\,k \\
        QCD $\pthat \in [15, 3000]\GeV$ & flat 55-75 & 100\,k \\
        QCD $\pthat \in [3000, 7000]\GeV$ & flat 55--75 & 100\,k \\
        \ztautau all-hadronic & flat 55--75 & 100\,k \\
        \hline
        single $\Pe$ flat $\ptmomentum\in[1,1000]\GeV$ & no PU & 10\,k \\
        single $\Pgm$ log-flat $\ptmomentum\in[0.1,2000]\GeV$ & no PU & 10k \\
        single $\PGpz$ flat $\ptmomentum\in[0,1000]\GeV$ & no PU & 10\,k \\
        single $\PGppm$ flat $\ptmomentum\in[0.7,1000]\GeV$ & no PU & 10\,k \\
        single $\PGt$ flat $\ptmomentum\in[1,1000]\GeV$ & no PU & 10\,k \\
        single $\PGg$ flat $\ptmomentum\in[1,1000]\GeV$ & no PU & 10\,k \\
        single $\Pp$ flat $\ptmomentum\in[0.7,1000]\GeV$ & no PU & 10\,k \\
        single $\Pn$ flat $\ptmomentum\in[0.7,1000]\GeV$ & no PU & 10\,k \\
        \hline
    \end{tabular}
    \caption{\acs{MC} simulation samples used for optimizing the \acs{MLPF} model. The kinematic quantity $\pthat$ is computed as the scalar sum of the outgoing generator-level partons.}
    \label{tab:samples}
\end{table}

\section{Event loss scans}
Since the end goal of \ac{PF} is global event reconstruction, we expect that \ac{MLPF}, in addition to reconstructing \ac{PF}-candidates, is able to reconstruct event-level quantities with high accuracy, such as those related to jets and the missing transverse momentum.
For each event, jets are clustered from reconstructed 
 or generator-level particles using the anti-\kt algorithm~\cite{Cacciari:2008gp, Cacciari:2011ma} with a distance parameter of 0.4 (AK4 jets).
The missing transverse momentum vector \ptvecmiss is computed as the negative vector sum of the transverse momenta of all the PF candidates in an event, and its magnitude is denoted as \ptmiss~\cite{CMS:2019ctu}.
The \ac{PUPPI} algorithm~\cite{Sirunyan:2020foa,Bertolini:2014bba} is applied to reduce the pileup dependence of the jet and \ptvecmiss observables~\cite{CMS:2019ctu}.

While the basic \ac{MLPF} algorithm incorporates a per-particle loss, it is interesting to study if including additional terms in the loss function can lead to better event-level reconstruction.
In general, the simple approach of clustering the reconstructed particles to jets, and comparing the resulting reconstructed jets to the generator-level jets is not practical, as fast, differentiable and GPU-optimized jet clustering algorithms are not yet available.
We have tested the following proxy event-level loss terms:
\begin{enumerate}
    \item Baseline: only the basic \ac{MLPF} per-particle classification and regression loss
    \item Sliced Wasserstein distance: compute a $\ptmomentum$-weighted mass transportation cost in the $[\eta, \sin{\phi}, \cos{\phi}]$ space between the set of reconstructed and target particles, approximating the metric through random projections (slicing)
    \item Generator-level jet $\log{\cosh}$: assuming that the reconstructed particles are clustered to the same jets as the target (\ac{MLPF} truth) particles, compute the effective reconstructed jet \ptmomentum values, and compare these to the generator-level jet \ptmomentum values using a $\log\cosh$ loss
    \item Missing transverse momentum: compute the reconstructed \ptmiss from the reconstructed particle candidates, and compare it with a mean squared error loss term with the target \ptmiss
\end{enumerate}

\begin{figure}[ht]
    \centering
    \includegraphics[width=0.49\linewidth]{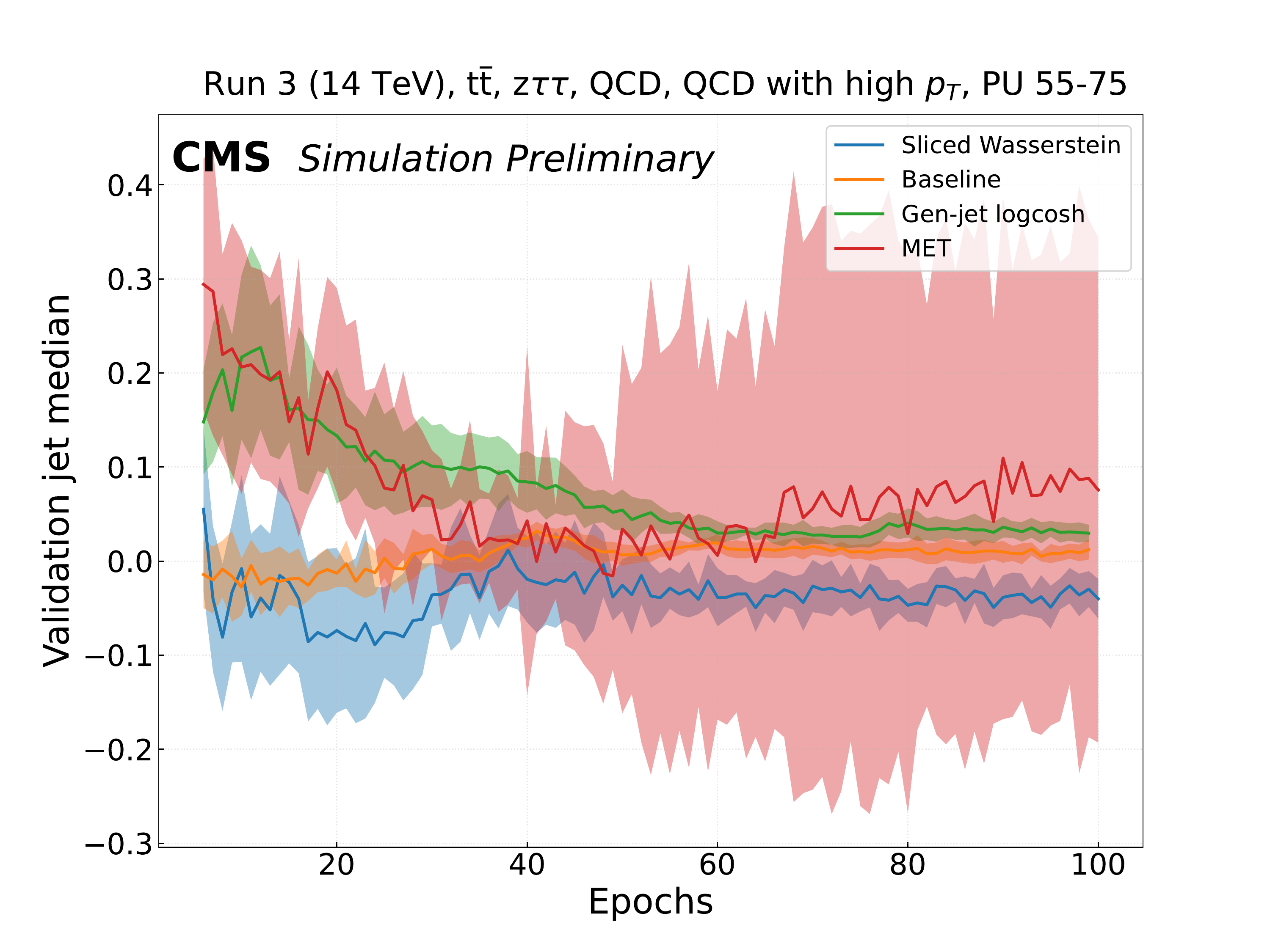}
    \includegraphics[width=0.49\linewidth]{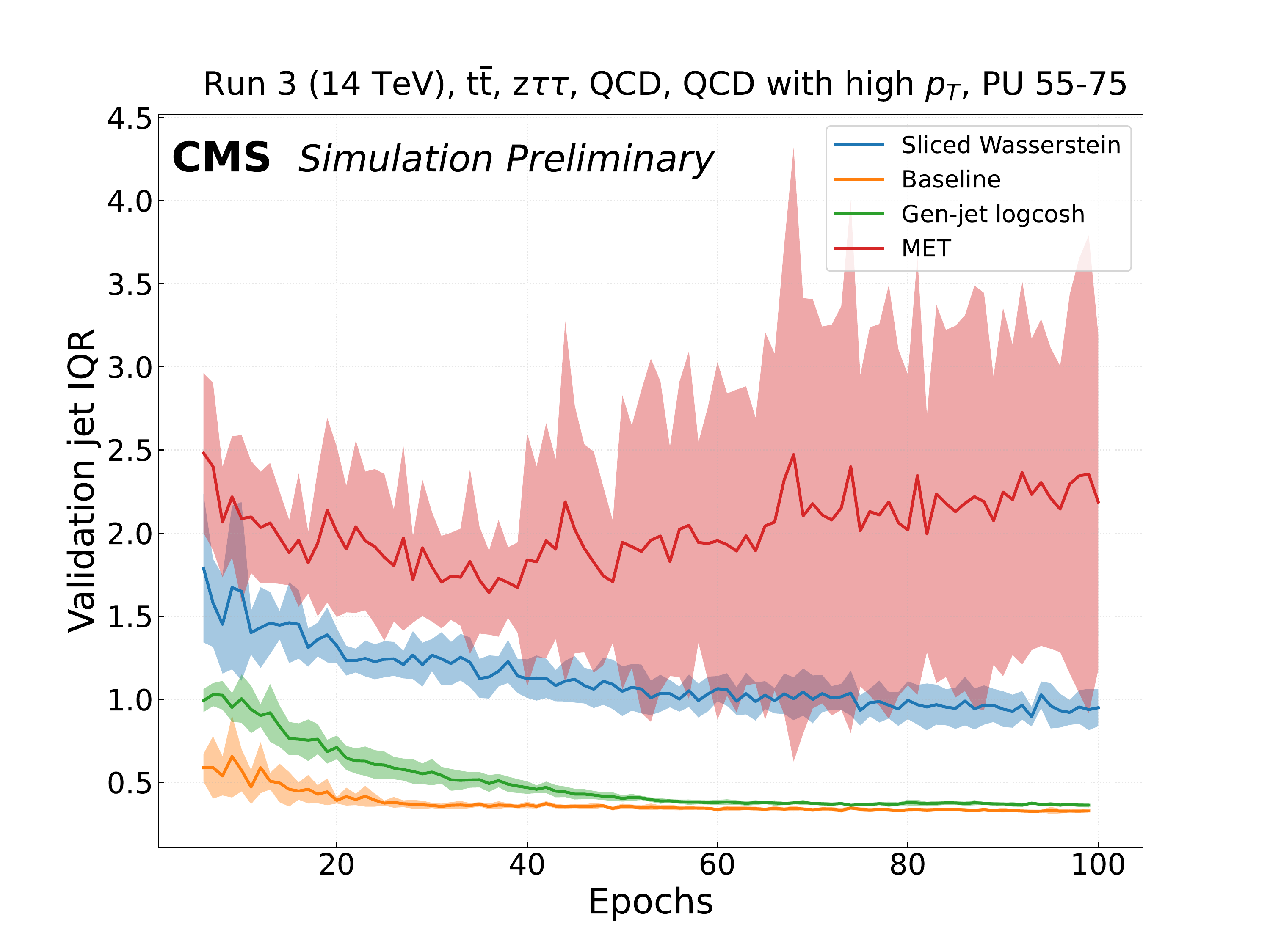}\\
    \caption{Comparison of the jet response median (left) and interquartile range (right) using different event-level loss terms. Solid lines show the mean of 10 trainings with identical configurations and shaded regions show the mean plus and minus one standard deviation of the 10 trainings.}
    \label{fig:event_loss}
\end{figure}

Comparisons are shown in \autoref{fig:event_loss}.
We monitor the median and interquartile range of the jet response distribution, defined such that the optimal values for both are zero.
We find that none of the tested loss functions improve the jet-level physics quantities that we observe during the training process, with the baseline, i.e. no additional event loss term added to the particle-level loss, performing the best.

\section{Event-level validation in CMSSW}
The \ac{MLPF} algorithm is integrated with \ac{CMSSW} through an optional (default disabled) switch, using \ONNXRUNTIME for inference.
The purpose of this validation is to test \ac{MLPF} in a real integration, completely independently of the \ac{ML} training data format and software framework. 
We demonstrate the results of enabling \acs{MLPF} in CMSSW reconstruction instead of standard \acs{PF} by validating on the following samples, which were not used in training: QCD multijet\footnote{The background from standard model events composed uniquely of jets produced through the strong interaction is referred to as quantum chromodynamics (QCD) multijet events.} with $\pthat \in [15, 3000]\GeV$ (46\,k events)\footnote{\verb|/RelValQCD_FlatPt_15_3000HS_14/CMSSW_12_3_0_pre6-PU_123X_mcRun3_2021_realistic_v11-v1/GEN-SIM-DIGI-RAW|} and top quark-antiquark pairs, or \ttbar, (8\,k events)\footnote{\verb|/RelValTTbar_14TeV/CMSSW_12_3_0_pre6-PU_123X_mcRun3_2021_realistic_v11-v1/GEN-SIM-DIGI-RAW|}.

The following results are event-level validation extracted directly from the MINIAOD event data.
We also test the performance of MLPF on the particle-level quantities, which are presented in full in the Detector Performance note~\cite{CMS-DP-2022-061}.

\subsection{Jets}

\begin{figure}[ht]
    \centering
    \includegraphics[width=0.35\linewidth]{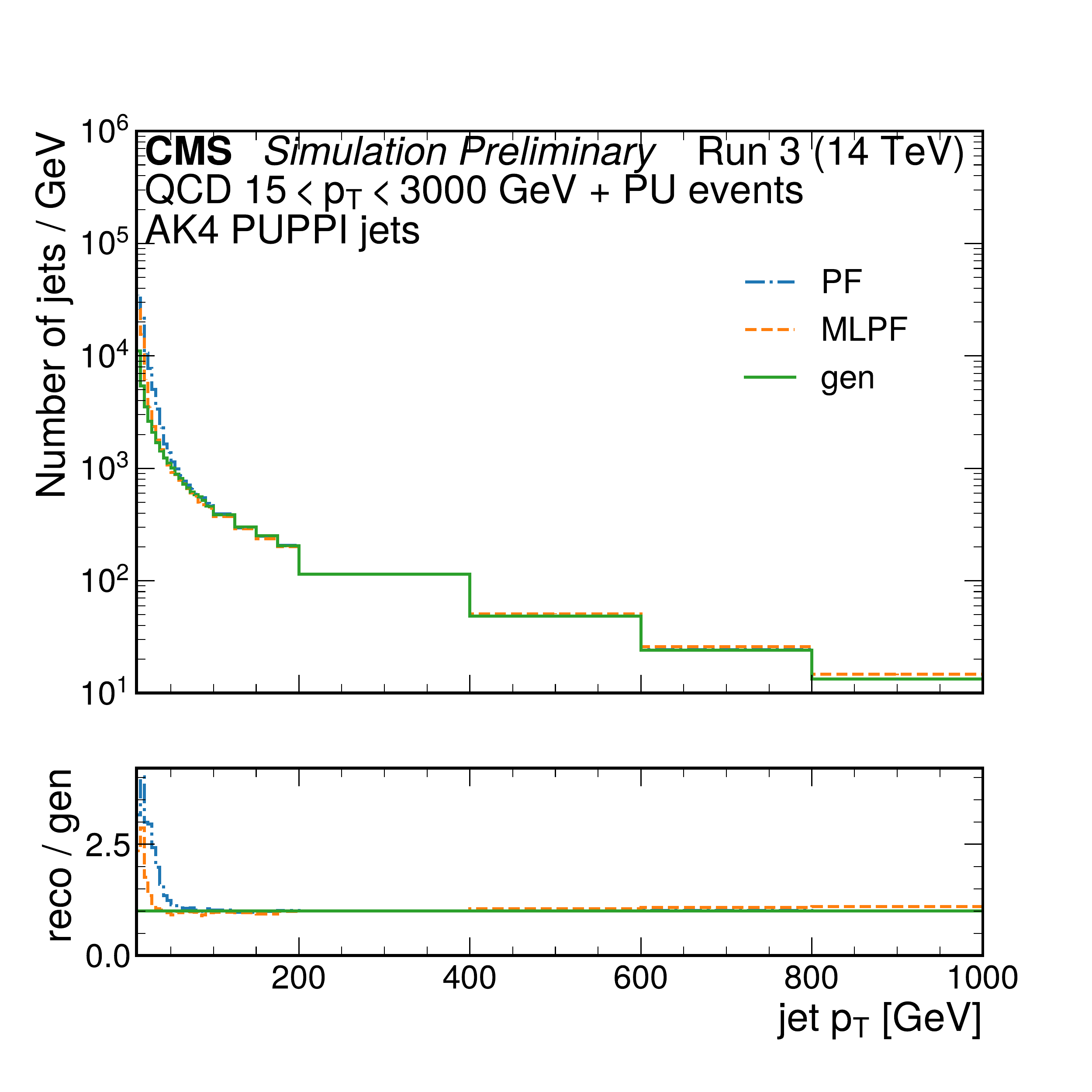}
    \includegraphics[width=0.35\linewidth]{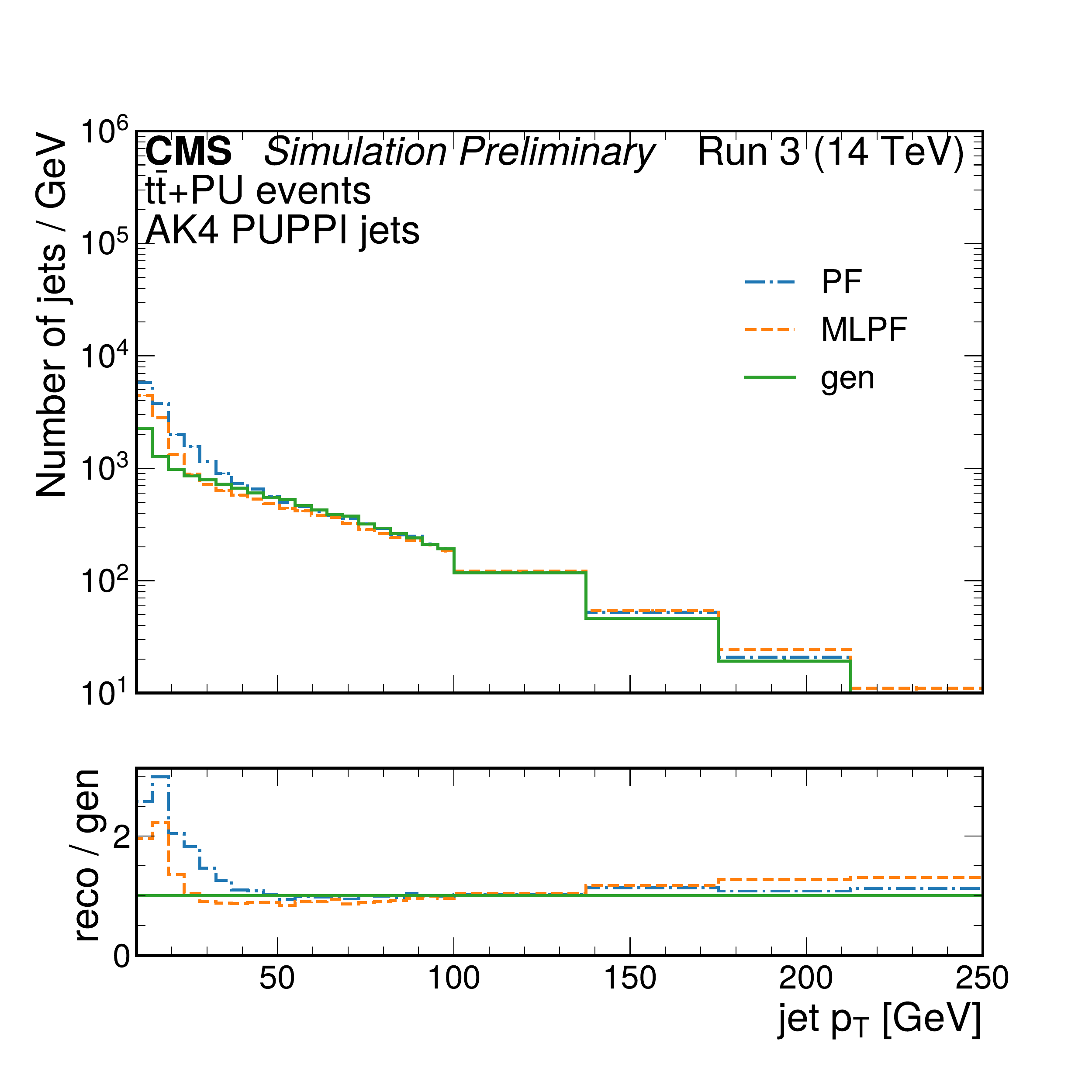}\\
    \caption{Reconstructed \acs{PUPPI} jet $\ptmomentum$ distributions with \acs{PF} and \acs{MLPF}, compared with the true generator-level jet distribution in QCD (left) and \ttbar (right) samples with pileup, reconstructed in CMSSW.}
    \label{fig:event_level_jet_pt}
\end{figure}

The comparison of generator-level and reconstructed jet distributions from \ac{PF} and \ac{MLPF} is shown on \autoref{fig:event_level_jet_pt}.
The jet distributions between \ac{PF} and \ac{MLPF} are broadly compatible, highlighting the jet reconstruction performance of the purely \ac{ML}-based \ac{MLPF} reconstruction that was trained on a generator/simulator particle level target. 

\subsection{Missing traverse momentum}
\begin{figure}[ht]
    \centering
    \includegraphics[width=0.35\linewidth]{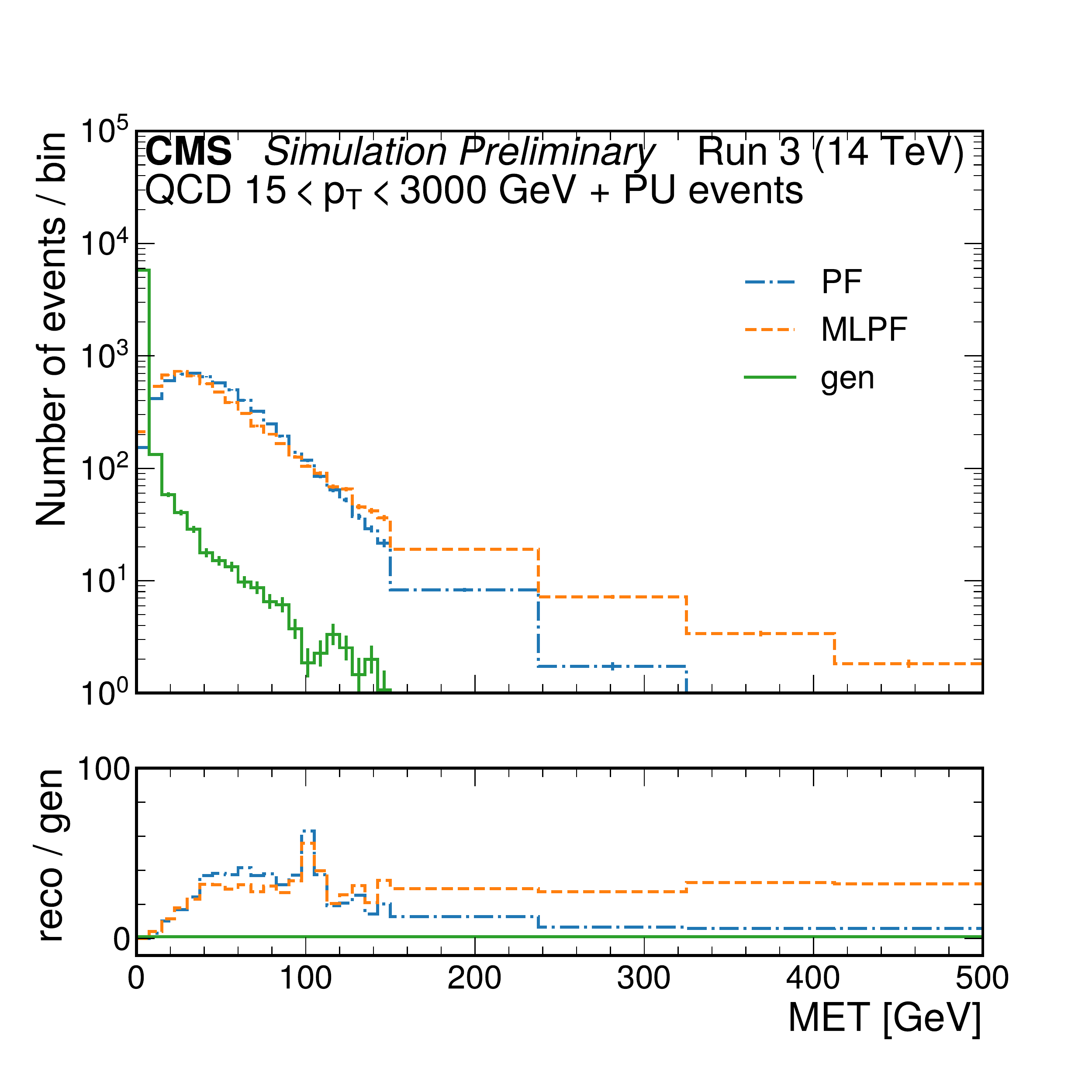}
    \includegraphics[width=0.35\linewidth]{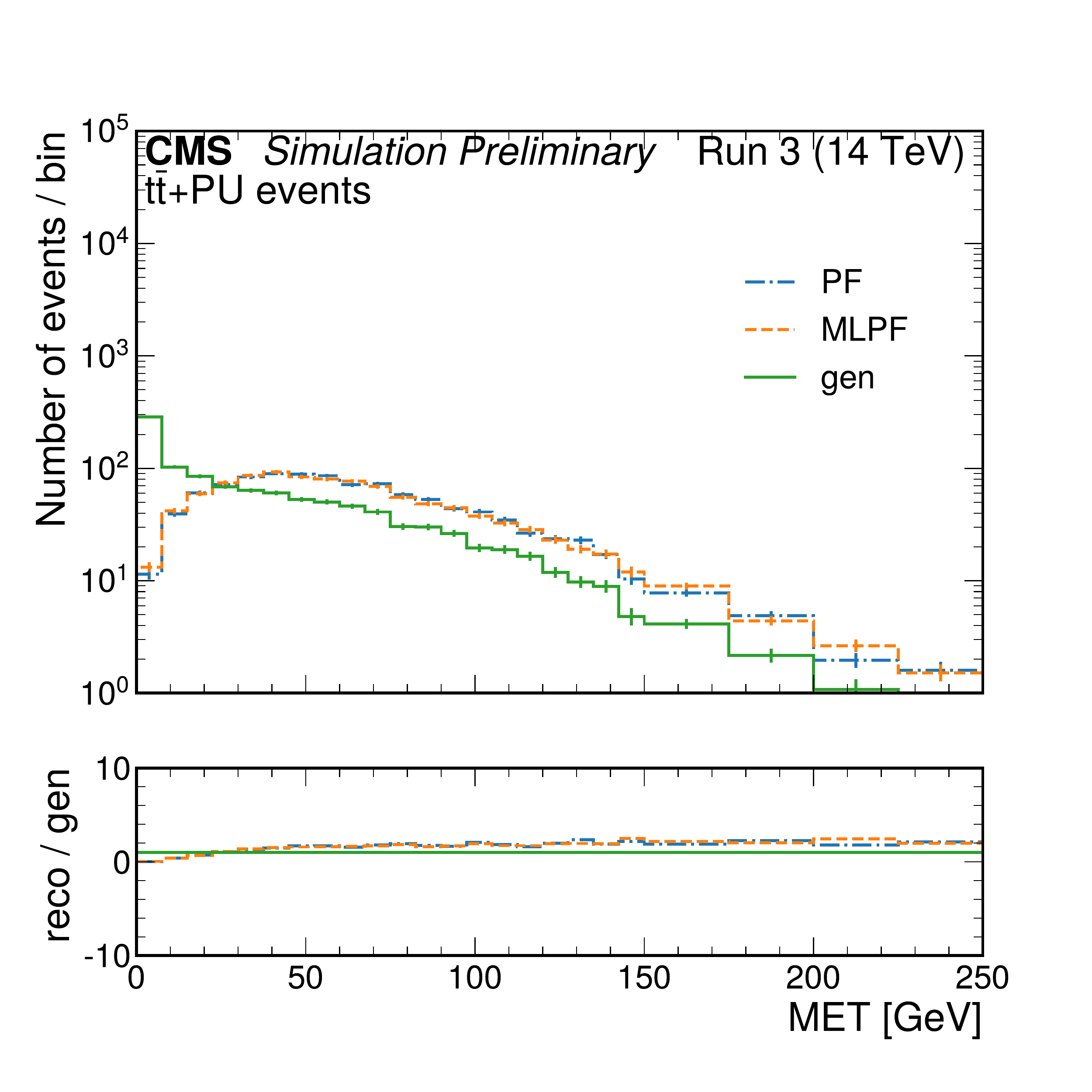}\\
    \caption{Reconstructed \acs{PUPPI} \ptmiss distributions with \acs{PF} and \acs{MLPF}, compared with the true generator-level \ptmiss distribution in QCD (left) and \ttbar (right) samples with pileup, reconstructed in CMSSW.}
    \label{fig:event_level_met_pt}
\end{figure}

The comparison of generator-level \ptmiss to reconstructed \acs{PUPPI} \ptmiss distributions from \acs{PF} and \acs{MLPF} is shown on \autoref{fig:event_level_met_pt}.
For the QCD sample in particular, we observe high-\ptmomentum tails in both \ac{PF} and \ac{MLPF}, which are more prominent in \ac{MLPF}.
Work is currently ongoing to reduce this effect through improved data samples and additional training.

\acresetall
\section{Conclusion}
We have defined a new generator/simulator-level \ac{MLPF} training target which enables the \ac{MLPF}-based reconstruction, in principle, to surpass the baseline PF algorithm reconstruction.
We presented comparisons between PF and MLPF on the particle-level, where the \ac{MLPF} performance is comparable to the rule-based \ac{PF}.
As a particular quantitative improvement, we find that the neutral hadron \ptmomentum response width is improved by about a factor of 2, while the \ac{MLPF} reconstruction is fully efficient at a lower fake rate than the baseline \ac{PF} for calorimeter clusters with $E > 10\GeV$.
We also compare the performance on the event level, directly in \ac{CMSSW}, where we find broadly comparable performance for jets, while the \ptmiss performance is affected by slightly larger tails in \ac{MLPF} than in \ac{PF} which we plan to address in a follow-up.
To improve the reconstruction of the event-level quantities further, we have explored the use of additional event-level loss terms, however, we find that in this case, the baseline particle-level loss performs the best.

The next steps include further work on the training datasets and on the \ac{ML} modelling and optimization, to further improve the reconstruction of complex event-based quantities.
It is also worthwhile to explore the application of explainable AI techniques on MLPF, since a common disadvantage of ML-based algorithms is a lack of interpretability.
In previous work we have explored such techniques in order to gain insight into the model's decision-making~\cite{Mokhtar:2021bkf}.
It is interesting to revisit such studies again, now that we have a well-defined MLPF truth.
Additionally, we also plan to rerun our hypertuning on HPC resources in light of the new dataset~\cite{Wulff:2022qep}.

\section{Acknowledgements}
We thank our colleagues in the CMS Collaboration, especially in the Particle Flow, Physics Performance and Dataset, Offline and Computing, and Machine Learning groups, in particular Kenichi Hatakeyama, Lindsey Gray, Jan Kieseler, Danilo Piparo, Gregor Kasieczka, Salvatore Rappoccio, Kaori Maeshima, Kenneth Long, and Juska Pekkanen for helpful feedback in the course of this work.
JP was supported by the Mobilitas Pluss Grants MOBTP187, PRG780 and MOBTT86 of the Estonian Research Council.
JD and FM were supported by DOE Award Nos. DE-SC0021187 and DE-SC0021396 (FAIR4HEP) and NSF Cooperative Agreement OAC-2117997 (A3D3).
FM was also supported by a UCSD HDSI fellowship and an IRIS-HEP fellowship through NSF Cooperative Agreement OAC-1836650. 
EW was supported by the CoE RAISE Project which have received funding from the European Union's Horizon 2020 – Research and Innovation Framework Programme H2020-INFRAEDI-2019-1 under grant agreement no. 951733.
Access to GPUs was supported in part by NSF awards CNS-1730158, ACI-1540112, ACI-1541349, OAC-1826967, OAC-2112167, CNS-2100237, CNS-2120019, the University of California Office of the President, and the University of California San Diego's California Institute for Telecommunications and Information Technology/Qualcomm Institute. 
Thanks to CENIC for the 100\,Gbps networks.

\printbibliography

\end{document}